\documentclass[11pt]{article}
\usepackage{graphicx}
\usepackage[margin=1.25in]{geometry}
\usepackage[usenames,dvipsnames]{color}
\usepackage{url}
\usepackage[colorlinks = true,
            linkcolor = blue,
            urlcolor  = blue,
            citecolor = blue,
            anchorcolor = blue]{hyperref}

\usepackage{lineno,hyperref,color}
\modulolinenumbers[5]
\usepackage{todonotes}
\usepackage{ulem}
\usepackage{booktabs}
\usepackage{float}
\usepackage{wrapfig}

\usepackage{appendix}
\usepackage{pdfpages}
\usepackage{hyperref}
\usepackage{geometry}                
\usepackage[parfill]{parskip}    
\usepackage{graphicx, subfigure}
\usepackage{amssymb}
\usepackage{amsmath}
\usepackage{epstopdf}
\usepackage{array}
\usepackage{lineno}
\usepackage{scrextend}

\def\Title#1{\begin{center} {\LARGE #1 } \end{center}}
\def\Author#1{\begin{center}{ \sc #1} \end{center}}

\newenvironment{Abstract}{\begin{quotation} \begin{center}
                       ABSTRACT
     \end{center}\bigskip  }{\end{quotation}}

\def\beq{\begin{equation}}
\def\eeq#1{\label{#1}\end{equation}}
\def\eeqn{\end{equation}}

\newenvironment{Eqnarray}%
   {\arraycolsep 0.14em\begin{eqnarray}}{\end{eqnarray}}
\def\beqa{\begin{Eqnarray}}
\def\eeqa#1{\label{#1}\end{Eqnarray}}
\def\eeqan{\end{Eqnarray}}

\let\bar=\overbar

\def\lsim{\mathrel{\raise.3ex\hbox{$<$\kern-.75em\lower1ex\hbox{$\sim$}}}}
\def\gsim{\mathrel{\raise.3ex\hbox{$>$\kern-.75em\lower1ex\hbox{$\sim$}}}}

\def\del{\partial}
\def\Dslash{\not{\hbox{\kern-4pt $D$}}}
\def\dslash{\not{\hbox{\kern-2pt $\del$}}}
\def\pslash{\not{\hbox{\kern-2pt $p$}}}
\def\ETmiss{\not{\hbox{\kern-4pt $E$}}_T}

\def\Dlr{\mathrel{\raise1.5ex\hbox{$\leftrightarrow$\kern-1em\lower1.5ex\hbox{$D$}}}}

\def\MSB{{\bar{M \kern -2pt S}}}
\def\msb{{\bar{\scriptsize M \kern -1pt S}}}

\def\drb{{\bar{\scriptsize D \kern -1pt R}}}

\input{defs}

\newcommand\snowmass{\begin{center}\rule[-0.2in]{\hsize}{0.01in}\\\rule{\hsize}{0.01in}\\
\vskip 0.1in Submitted to the  Proceedings of the U.S. Community Study\\ 
on the Future of Particle Physics (Snowmass 2021)\\ 
\rule{\hsize}{0.01in}\\\rule[+0.2in]{\hsize}{0.01in} \end{center}}

\begin{document}


\Title{Enabling U.S. participation in Future Higgs Factories}

\bigskip
\Author{K.~Black$^{1}$,
K.~Bloom$^{2}$,
J.E.~Brau$^{3}$,
M.~Demarteau$^{4}$,
D.~Denisov$^{5}$,
D.~Elvira$^{6}$,
S.~Eno$^{7}$,
R.~Hirosky$^{8}$,
J.~Hirschauer$^{6}$,
R.~Lipton$^{6}$,
C.~Paus$^{9}$,
E.~Stern$^{6}$,
A.~White$^{10}$,
G.W.~Wilson$^{11}$}

\begin{center}
$^{1}$University of Wisconsin,
$^{2}$University of Nebraska-Lincoln,
$^{3}$University of Oregon,
$^{4}$Oak Ridge National Laboratory,
$^{5}$Brookhaven National Laboratory,
$^{6}$Fermi National Accelerator Laboratory,
$^{7}$University of Maryland, 
$^{8}$University of Virginia,
$^{9}$Massachusetts Institute of Technology,
$^{10}$University of Texas, Arlington,
$^{11}$University of Kansas
\end{center}


\medskip

\medskip

 \begin{Abstract}
\noindent
Exciting proposals for a new ``Higgs factory'' collider, 
aimed at the search for new physics and precision studies of particles and forces, especially
measurement of the Higgs boson couplings at the loop level, will be evaluated as part of the Snowmass process.  
Potential facilities include (among others) ILC, FCC-ee, C$^3$, CEPC, CLIC, muon collider and advanced accelerator concepts being investigated by Snowmass topical group AF6, potentially located in Asia, Europe, or the United States.
The European Strategy has endorsed an $e^+e^-$ Higgs factory as its highest priority after HL-LHC.
Much of the detector, software, and physics preparative studies needed for these machines is in common, and is currently being implemented
by physicists world-wide.
In this white paper for the 2021 Snowmass process
we look at current global activity on future Higgs factories and give examples of  investments that could be made in these common areas over the next five years to establish a leadership role for the U.S. in a future Higgs factory, wherever it is built.   
The U.S. high energy physics program confronts a number of challenges that a strong role in the study of the Higgs boson can address.  These include, in addition to the scientific results, maintaining leading roles in international partnerships, nurturing and advancing world-leading capabilities and expert resources, and maintaining and attracting talent. 
The international effort would benefit from increased U.S. participation, and the U.S., in turn, would maintain stature through the partnership.


\end{Abstract}

\snowmass

\def\thefootnote{\fnsymbol{footnote}}
\setcounter{footnote}{0}
%


\section{Introduction}
The United States has a long tradition of intellectual leadership in collider-based studies of fundamental physics. 
Our current leading contributions  to the physics program of the LHC, including  the discovery of the Higgs boson in 2012~\cite{higgscms,higgsatlas}, are built on a foundation that started more than ten years before the start of collisions.  The total pre-LHC investment
is even larger when investment in the Superconducting Super Collider (SSC) is included.
The SSC detector design and development, software design and development, and physics studies, led into leadership roles in LHC detectors, software, and physics analysis.
Investment directly in the SSC and LHC programs was concurrent with the Tevatron detector upgrades and data taking.

The world is coming to a consensus that a Higgs factory is the natural next highest priority major HEP collider beyond the HL-LHC~\cite{EuropeanStrag}.
Potential facilities include (among others) ILC, FCC-ee, C$^3$, CEPC, CLIC, muon collider and advanced accelerator concepts being investigated by Snowmass topical group AF6, potentially located in Asia, Europe, or the United States.
The European Strategy has endorsed an $e^+e^-$ Higgs factory as its highest priority after HL-LHC~\cite{EuropeanStrag}.
The exciting physics programs allowed by these facilities will be discussed in detail in other submissions to the Snowmass process.
Physicists world-wide are already laying the groundwork necessary for these future machines, through accelerator design, detector design, and physics sensitivity studies.  Their efforts are also becoming increasingly coherent through unified work {\it e.g. }on physics generators. 

The future of the U.S. high energy physics program depends on succeeding with a number of challenges.  Foremost are the scientific outcomes. In addition, U.S. international partnerships are a critical aspect, and maintaining leading roles within those partnerships are needed to ensure continued success in the future.  Within the diverse U.S. community there are many world-leading capabilities and expert resources that enable the program and need to be nurtured and advanced.  Successful outcomes depend on maintaining and attracting talent; this requires attractive prospects for on-going and future scientific opportunities.

U.S. physicists engagement in this global work, to have an appropriate impact, needs a substantial level of investment
 well in advance of the finalization of the accelerator and detector designs. This will ensure leadership in the process of designing the accelerator complex and the detectors. It should be 
similar in funding level to that of other participating countries.  
In this white paper, we give examples of areas where investment can be made that will be impactful regardless of the eventual details and location
of the facility.  
We concentrate on impact on 
accelerator software,
detector development, software and computing, and the physics program, as these are areas where this paper's authors have expertise.
We also compare our current levels of impact to those of other countries.

Such investment will also show the world that the U.S. is strongly committed to the exciting physics program enabled by Higgs factory experiments.  
By investing now, and contributing to the developing, integrated world community, we can ensure U.S. leadership
in the construction of a facility
that is able to commence operation before, or shortly after, the end of the HL-LHC program. The  expertise and strength of the
U.S. community will be central to the program.
By working together with our international partners, we give any potential host nation 
confidence that the non-host nations will deliver the needed contributions.




\section{Vision}
U.S. potential for participation in future Higgs factories could be revolutionized by having a small but full-time core group of  researchers working on Higgs factory research.  Working with the rest of the laboratory community and the community at U.S. universities, these dedicated personnel would enable the efficient participation by others on a part-time basis.  
Ideally a core at national  labs would be supplemented by a few full-time university-based research scientists, engineers and technical specialists.
This joint laboratory-university full-time team will maintain the long-term memory and mentoring of new younger people who would join the project, playing the role that many U.S. lab personnel and university-based research scientists played in early U.S. involvement in LHC. 
 In addition, work in the area of detector development will ensure the current expertise is passed down to a new generation.
 
 For maximum impact, all of this needs to be seamlessly incorporated into the existing international effort
 and to build faith in the world community that all participating nations are committed to a Higgs factory.

\section{Potential investment areas }

There are many areas where strategic investment now by the United States is necessary to ensure a strong U.S. role no matter which factory
is built.  While we discuss a limited set of examples here, { \bf these should be taken only as examples of areas where investment now will lead to more impactful U.S. participation in future facilities}.

\subsection{Accelerator modeling}

An essential area for U.S. investment for high impact on future Higgs factories is in accelerator simulation software and computing infrastructure to support design and construction of machines.  Such infrastructure and core knowledge can be used to study how to
overcome the challenges of producing and controlling beams of increasing intensity.  Accelerator design requires expert teams at labs and universities equipped with detailed beam modeling toolkits that simulate the collective physics effects that limit accelerator performance as well as commensurable computing infrastructure to perform the simulations. Development and validation of the software requires contributions from beam physics and software experts to ensure correct results and user-friendly interfaces.
 
U.S. effort in this area is currently small.  However, renewed effort in this area would be strongly welcomed by the international community and help ensure a Higgs factory start before the end of, or shortly after, the HL-LHC.

\subsection{Detector R\&D for Higgs-factory detectors}
In order for the U.S. to enjoy ``joint ownership" for a subdetector at a future collider, it is necessary that university  and laboratory groups  get in on the ground floor.  If we take the example of the pixel detectors, the pioneering work at FNAL and LBNL led to strong U.S. impact in the pixel design and construction.  As physicists transitioned from the Tevatron to LHC, they naturally found projects in the areas where the U.S. had already provided investment.  
There is an additional benefit to funding detector R\&D for Higgs factories:  as the HL-LHC upgrades are moving into the construction phase, work on practical R\&D will also allow students to get experience in the design and prototyping of detectors.

Calorimetry development for future Higgs factories is currently dominated by CALICE (for high granularity) and IDEA (for dual readout).  The U.S. is no longer supported for CALICE calorimeter prototype development, design, or testing, and also does not have funding for dual readout work.  
Both the CALICE collaboration and the members of the IDEA collaboration have frequently expressed a strong welcome for extended U.S. involvement.  U.S. efforts here would lead to a stronger international community  furthering calorimeter design.
The U.S. retains strong expertise in these areas from previous investments.
At U. Washington a dual readout concept was first proposed, recognizing the measurement of the two main components in a shower with multiple sampling media to improve hadron calorimetry~\cite{mockett}. This approach was subsequently implemented by the DREAM Collaboration~\cite{RevModPhys.90.025002}, with leadership  from Texas Tech.
The U.S. laboratories and Caltech have long advocated for homogeneous hadron calorimetry. 
The U.S. was also initially part of the group developing high granularity calorimetry (HGCAL), and institutions such as NIU, Iowa, U. Texas Arlington, Oregon, Argonne, Washington, Colorado, and SLAC played leading roles.  The U.S. (Oregon and Tennessee) also constructed the first silicon-tungsten luminosity monitor calorimeter, at the SLD experiment at SLAC~\cite{Berridge:1988ie}.


Another example of an area where increased U.S. involvement could lead to a strong impact is silicon detector development for tracking. 
Precision measurements are demanded by searches for small deviations from the Standard Model and imply a precise, low mass tracker. Having excellent momentum resolution for e.g. muons leads (close to linearly) to reduced uncertainties on Higgs mass measurements and decay-mode-independent Higgs identification via recoil mass measurement using the associated recoiling Z boson. The muon collider adds to the $e^+e^-$ constraints the additional challenges of 
large Beam Induced Backgrounds (BIB) and significant levels of radiation. In either case design and construction will be 
challenging, and we must be prepared to develop and adopt new technologies where appropriate.

A U.S. effort led by SLAC National Laboratory in collaboration with the University of Oregon, UT-Arlington and Oak Ridge National Laboratory is developing CMOS Monolithic Active Pixels (MAPs) for applications in tracking and electromagnetic sampling calorimetry for the linear collider~\cite{ilcx2021}.
Larger areas of silicon sensors are needed, several
hundred m$^{2}$, for the low mass trackers and sampling calorimetry.
Trackers require multiple layers, large radii, and micron scale resolution. The CMOS MAPs application present a promising approach, in which silicon diodes and their readout are combined in the same pixels, and fabricated in a standard CMOS process.
CMOS MAPs sensors have several advantages over traditional hybrid technologies with sensors bonded to readout ASICs. These include  the sensor/front-end electronics integration, reduced capacitance and resulting noise, lowered signal to noise permitting thinner sensing thickness, very fine readout pitch, and standardized commercial production.  A number of challenges to the development include the limited reticle size (about 2 cm by 2 cm) and required stitching of reticles to 10 cm by 10 cm.  ALICE is addressing many challenges with its Inner Tracking System Upgrade based on ALPIDE, but others remain. This project aims to address them, with a prototype that solves powering and readout issues of large numbers of sensors, achieving finer granularity, higher resolution, and reduced mass budget, at much lower cost. 

Low Gain Avalanche Diodes (LGADs) offer the possibility of combined fast timing, low mass, and precision spatial resolution.  3D integration is now a standard industry technology, offering dense, heterogeneous, multilayer integration of sensors and electronics. Advanced process nodes are available as well, lowering power and increasing the density of integration. Mechanical supports are getting lighter, with ${\rm CO_2}$ cooling, carbon fiber, and carbon forms included in current tracker designs. 

The U.S. HEP community has had a long history in designing and building tracking detector systems for colliding beam experiments. 
The CDF CTC drift system and SVX silicon systems were pioneering detectors for hadron colliders.  This was followed by the D0 
SMT tracker and the CDF and D0 designs for Tevatron Run 2b detectors.  Run 2b was cancelled, but the Run2b design work was the 
basis of the D0 Layer 0 detector, a predecessor of the ATLAS insertable B layer; and for the ATLAS HL-LHC tracker mechanical 
designs. US groups developed a complete silicon-based tracker design for SiD detector at ILC, including the sensors, mechanical support 
structures, air cooling and electronics. The US also developed first tracking concepts for a muon collider, based on the SiD design. 

The U.S. also has had leadership in sensor development. LBNL and Brookhaven have led the way with in-house fabrication facilities.  This work included technical 
development of low-leakage, radiation hard sensors and the development of "3D" radiation hard sensors based on deep etching techniques. 
This work spun off the deep depletion CCD, used for dark matter studies and infrared astronomy and the extremely low noise "skipper" CCD. 
These groups are actively pursuing R\&D in new and emerging technologies, including AC-coupled LGADs for fast timing. 

There was also pioneering work in detector electronics, including the Microplex chip, the first ASIC designed for detector readout, 
followed by the SVX family of chips and adaptations of the architectures for $e^+e^-$ and hadron colliders. Early work on pixel detectors was led by SLAC, demonstrating bump bonding and "data push" designs date back to the early 1990's.  LBNL has been a leader in pixel readout chips and 
 hybridization. The first demonstration of 3D integration of sensors and electronics was by a FNAL-led consortium which developed an 
 integrated three-layer detector/readout stack with the two IC layers having a total thickness of 35 microns interconnected by through-silicon-vias. 
 
Trackers are integrated systems, where tradeoffs must be made between mass, power, speed, resolution, and processing. Broad experience 
is crucial to successful design and assembly. One generation often leads directly to the next, with lessons learned and opportunities 
presented by R\&D recognized. The U.S. experience and unique resources of the 
combined university/national laboratories complex must not be lost. This is not only possible, but likely, if  the U.S. does not engage 
wholeheartedly in new initiatives. Early engagement often translates into leadership and that leadership is crucial for a healthy 
enterprise.


\subsection{Software needs for physics, detector, and accelerator studies}

The software and computing needs of analysis efforts to make the physics case for Higgs factories, as well as the needs for simulations of potential detectors   to demonstrate their feasibility
need to be addressed without delay if the international community is serious about a Higgs factory operating before the end, or immediately after the completion of the HL-LHC physics program. Teams with expertise on detector modeling tools focused on future colliders need to be strengthened and provided with resources within high-energy physics laboratories and university groups. 

We know that studies for Higgs factories will demand significant effort on software infrastructure, simulation tools, and reconstruction algorithms capable of modeling hard collisions at the required energies. These efforts include software improvements and physics enhancements to event generators and to the Geant4 detector simulation toolkit, in order to enable detailed descriptions of the complex geometries of future detectors, accurate modeling of hard collisions and physics interactions inside the detectors, as well as to extract all the physics information delivered by novel detector technology and features. Even if computing demands of Higgs factories were smaller than those of HL-LHC, software packages need to be re-engineered to incorporate modern techniques, such as AI, and run on new computing platforms, including super-computing facilities requiring efficient use of hardware accelerators.

 The U.S. program needs a long-term commitment to build expertise through new hires and training, given that the development, maintenance and support associated with the above-mentioned tasks require skills and expertise which are scarce and in high demand. Investment should not be limited to the software tools themselves but also cover the expertise to integrate physics generators and detector configurations within the experimental frameworks. This includes efforts to tune and validate generated events, use high-level languages (such as DD4hep and Key4hep~\cite{Key4hep}) to describe detector geometries, optimize navigation, magnetic field, and physics options within Geant4, and model pileup backgrounds and readout electronics, and to assist the detector development community with the use of these tools.
 
  An important area of expertise to expand in the US is contributing to
complete reconstruction of events. Physics and detector
performance studies use sophisticated algorithms that seek to
reconstruct all physics objects using information from all relevant
sub-detector systems in an integrated way within a full collider event.
These are of critical importance for evaluating and improving the
overall collider detector design, and are needed to assess in a holistic
way the pros and cons of different sub-detector approaches, and can be
beneficial for understanding the potential of specific detector R\&D
opportunities in the Higgs factory environment.
Along these lines, it would be particularly useful to develop a
framework for evaluating the physics potential and detector performance of
various Higgs factory detector concepts, potentially across different
accelerator facilities.
 
 
 The U.S. HEP community has not succeeded in crafting a long-term coherent funding model to support physics generators, and has reduced detector simulation direct investment, e.g., Geant4 development, to a modest contribution to R\&D for new computing architectures. Continuity and predictability are essential to build competent and productive teams to provide software and computing support for future collider studies, thus enabling the US to play a leading role within the international community.

\subsection{Physics analysis preparations}
Involvement in physics reach studies and analysis preparation not only helps in the planning, design, and optimization of the physics program and experiments, it can provide opportunities for young faculty, postdocs, and graduate students to produce limited authorship papers that are useful in furthering their careers.  A ``school" training program  similar to those provided by the ATLAS and CMS collaborations ({\it e.g.} the CMSDAS at the LPC), organized by the core full-time community, would help them engage part time, providing full time experts in their own time zone and lowering travel costs.
Computing resources needed to support these studies could be provided through the Open Science Grid, with some assistance from their staff in deploying the necessary software.

\subsection{Computing}

Computing has turned out to drive a major component of the cost of LHC experiment operations.  The complexity of LHC collision events and the large trigger rates drive these costs, as they lead to demands for millions of processors and petabytes of disk storage to process and store both detector and simulated events.  These demands have been satisfied through the deployment of a global network of computing centers that share the load of data processing and storage.  The cost for these centers has been borne by the nations that deploy them, typically at a level that scales with their headcount on a given experiment.  Strong U.S. participation in a Higgs factory experiment will likely necessitate an equally strong need to contribute both financially and intellectually to the computing for that experiment.  The cost of the deployment and operations of U.S. computing facilities for the LHC is about a quarter of the total U.S. operations cost of the experiments, even with the institutes operating the facilities providing the infrastructure, power, and cooling for them.  There are additional costs for developing and maintaining the ``software" parts of the computing, namely the data management and workflow management infrastructure.

The computing demands of a Higgs factory are expected to be on the same scale as those of the LHC~\cite{miyamoto_akiya_2021_4659567}.  Because of the small lepton interaction cross section, event rates will be smaller, and in the absence of strong interactions and with a low number of multiple interactions the collision events should be simpler to interpret.  However, sophisticated detectors that provide rich information on particle interactions could require compute-intensive simulation and reconstruction algorithms, and if current trends continue physicists will be more and more interested in analyzing the data with compute-intensive artificial intelligence and machine learning techniques.  
Furthermore, this provides an opportunity to devise a computing infrastructure very different than that needed for the LHC (or the HL-LHC), which could allow for a different and better user experience.

It is not too early to begin to specify a computing model for a Higgs factory.  Efforts to develop computing models for the LHC started in the 1990s, well before first collisions, with the MONARC project completing around 2000~\cite{MONARC}.  MONARC made use of about 200 person-months of effort, the equivalent of about 5~FTE over three years.  A similar scale effort could make significant progress on a Higgs factory computing model.  This could begin with a widely agreed-upon specification of the parameters that drive the model, such as event sizes and processing times, and then exploration of how those parameters drive requirements on computing facilities (processors, disk, tape) and the wide-area networks that connect them.  This would result in some preliminary cost estimates for Higgs factory computing, which would be necessary for any estimate of the true cost of experiment operations.

\subsection{Complementarity of U.S. laboratories and university groups}
The United States is fortunate to have both strong federally supported  laboratories
 and a wide variety of strong university groups.  For long range planning, these complement each other.
Support for university groups plays a particularly important role in
training outstanding young scientists in the field. Young scientists are very interested in what comes after the HL-LHC and have the state-of-the-art technical talents to perform detailed studies of physics measurements, and maintain a simple software framework to encourage new participants. However, to keep up on an experiment like ATLAS or CMS is already more than a full-time-job. A strong full time core group at labs can actuate this eager community and help make the Higgs factory a reality.
It would also benefit the U.S. program to restore some of the university-based research scientists.  These scientists have more flexibility regarding travel than lab-based researchers, can be less expensive, and have traditionally been leaders in long-term preparations.

\section{Needed investment size}

For a Higgs factory on a desirable timescale, a substantial investment is needed now.  In this section, we look at the level of funding that allowed the U.S. to have a strong impact on ILC detector design prior to 2008, and also at funding of Higgs factory efforts elsewhere in the world. Also, since in many cases detailed information is not available, we look at heuristics like talks given at conferences as a measure of engagement.

\subsection{Historic U.S. investment in Higgs factory accelerators }

During the period 2005 -- 2008, the DOE supported a broad range of activities by the Global Design Effort to develop the International Linear Collider, at the time the only Higgs factory under serious consideration.  At its peak, the funding was at the level of \$30M per year, with about 90\% devoted to the design of the facility and R\&D on critial subsystems.  The remaining 10\% was used for detector R\&D, support for the GDE management, and an international ILC school for young physicists.   Similar levels of funding and in-kind contributions were provided by Europe and Japan.  This funding was critical for completing the 2007 Reference Design Report and the subsequent Technical Design Report in 2013 that established the ILC as a mature project, ready for project initiation.

\subsection{Historic U.S. investment in Higgs factory detectors and physics program}

Detector development for ILC was funded from fiscal years 2005 to 2008~\cite{osti_1074469}, and was about \$3.4M in total from DOE
and about \$1.1M from NSF to the universities and a coordinated \$0.7M to the labs who were working
with the universities on specific projects.  It was during this time that the U.S. had a large impact on high granularity calorimetry through funding of NIU, Iowa, U. Texas Arlington, Oregon, Argonne, Washington, Colorado and SLAC.  The U.S. also had a leading role in thin silicon sensors via Purdue, Oregon and SLAC (now led by members of the ALICE collaboration in Europe).

\subsection{Dedicated Higgs factory funding outside of the U.S.}
In this section, we look at current investment in the Higgs factories outside of the United States.   

One thing to note when estimating the appropriate size of a potential U.S. contribution:  the population of the U.S. is 62\% that of the sum of the countries in the CERN member states.
Currently employees of U.S. institutions make up about 20\% of each of the ATLAS and CMS collaborations.  The U.S. population is 2.6 times that of Japan.
The GDP of the U.S. is  1.1 times that of the CERN member states and 4 times that of Japan.

Also note that this investment is being made simultaneously with the operation of the LHC and the construction of HL-LHC and its detector upgrades.

\subsubsection{Investment via CERN funding}
Full-time dedicated staff are essential for the establishment of a new facility and a new physics program.  Typically, the host laboratory  provides a substantial
portion of the dedicated staff.  For European nations, a common host laboratory is CERN.
Based on the outcome of the European strategy~\cite{EuropeanStrag}, CERN (and via CERN its member states) has decided to make a large investment in a future Higgs factory (see Figure 6 in Ref.~\cite{cernmed}).
Details are shown in Table~\ref{tab:cernbud22}.


\begin{table}
\begin{tabular}{ m{6.0cm}    m{1.5 cm}  m{1.5cm} m{1.5cm}} \hline
    name                                                     & personnel (MCHF) & Materials (MCHF) & Total (MCHF) \\ \hline
    High-field superconducting accelerator magnets (HFM) R\&D & 19.9 & 73.8 & 93.7\\
    Linear collider &  19 & 14.4 & 33.4 \\
    FCC &  34.3 & 85.6 & 119.9 \\
    Muon collider & 5 & 5 & 10 \\ \hline
    \end{tabular}
    \caption{Information on CERN "medium term plan" funding for 5 years}
    \label{tab:cernbud22}
\end{table}

\subsubsection{Italy}
Italy has also made public its plan to invest heavily in future Higgs factories.  INFN in its report for the next three year plan~\cite{italypiano} considers increasing the yearly support for 
Muon Collider and FCC R\&D from 1\% to 2.8\% of the CSN1 (the Italian funding committee for experiments at particle accelerators) budget of 20 M\texteuro, as indicated by its International Evaluation Committee. Funds for FCC in 2021 have been 234 k\texteuro\ and in 2022 this has grown to 345 k\texteuro. Additional INFN support for  dual readout calorimetry for about 800 k\texteuro\  in 3 years, starting in 2022 with 171 k\texteuro, is provided by CSN5 (the Italian technology R\&D funding committee).  
CSN5 also extended support for the pixel detector R\&D called ARCADIA for another year in 2022 with 185  k\texteuro~\cite{bedeschi}.

INFN leads two grants related
to Higgs factories: {\it AIDAinnova} and {\it Euro-Labs}.

{\it AIDAinnova}
 is an approved EU grant of 10 M\texteuro\ for a duration of 4 years. The project involves R\&D on basically all the different types of detectors in use or planned for HEP, with a strong emphasis on future experiments at future colliders (FCC-ee, ILC, CEPC, etc.). The project started in April 2021 and will therefore run until the end of March 2025. The project involves 45 beneficiaries, subdivided in 35  academic institutes and 10 Companies, plus 10 associated partners. AIDAinnova adopts a co-financing scheme in which each beneficiary co-finances the project. The total funding of AIDAinnova reaches about 25 M\texteuro.
Roughly a bit more than 30\% of the budget is reserved for the R\&D for future Higgs factories and Italy's share is a bit more than 3 M\texteuro.

{\it Euro-Labs}
 is an accepted EU grant for accessing test beams and irradiation facilities. It is a large project, submitted in conjunction between Nuclear Physics, Accelerator physics and Detectors for HEP. More than 30 European Research Infrastructures are included and accessible through Euro-Labs. The funding amount is 14.5 M\texteuro\ also over a timespan of 4 years. The starting date of the project will be September 2022.
This funding will pay for travel expenses to participate to test beams. Detectors for HEP will receive more than 3 M\texteuro\ over 4 years. 25 beneficiaries and 8 partners encompassing several European countries are involved in the project.

\subsubsection{S. Korea}
S. Korea is another nation that is actively pursuing leadership in a future Higgs factory.  R\&D for
dual readout calorimetry for future Higgs factories of  \$2M over 5 years~\cite{Skorea} has been allocated.

\section{Metrics on current U.S. impact on future Higgs factories}
Since funding numbers are not yet available from many countries, we look at other measures of their engagement.
There are several potential heuristics.
For example, one could look at the fraction of leadership positions in an official hierarchy.  
The intellectual impact of an individual in a small management team, however, is difficult to measure.
Perhaps more useful is the fraction of talks at workshops by country of affiliation, as these talks
generally represent cutting edge results and are solicited from the leaders in each area.

Data used in this analysis can be found at \href{
    https://docs.google.com/spreadsheets/d/15Ea51sFbuNILqeieF8dVOV5txJXa8N-KgyovrpN_XwA/edit?usp=sharing
}{this link}.

\begin{figure}
\centering
\resizebox{0.48\textwidth}{!}{\includegraphics{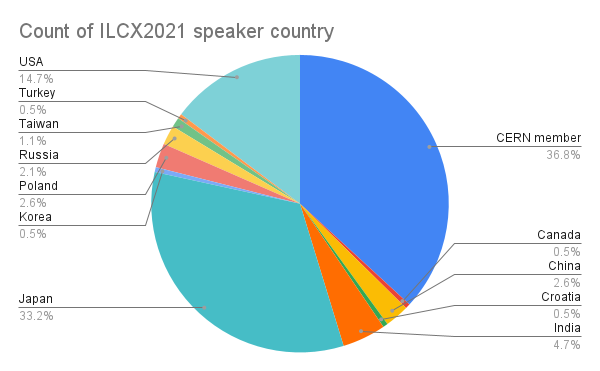}}
\resizebox{0.48\textwidth}{!}{\includegraphics{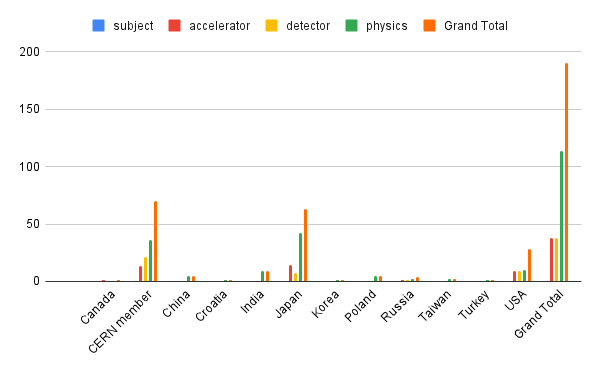}}
\caption{ 
for the \href{https://agenda.linearcollider.org/event/9211/}{ILCX2021 conference}  
[left] talks by country 
[right]   talks by subject
}\label{fig:ilcx2021}
\end{figure}

The \href{https://agenda.linearcollider.org/event/9211/}{ILCX2021 conference} was held fully online in October, 2021.  Figure~\ref{fig:ilcx2021} shows the nation of the speaker's affiliation for all talks (left) and by several talk categories (physics, accelerator, detector development) (right).

\begin{figure}
\centering
\resizebox{0.48\textwidth}{!}{\includegraphics{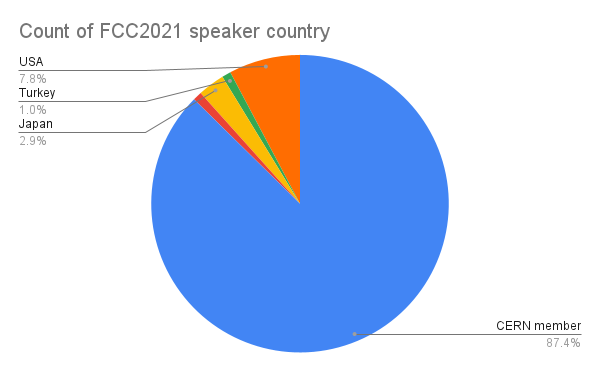}}
\resizebox{0.48\textwidth}{!}{\includegraphics{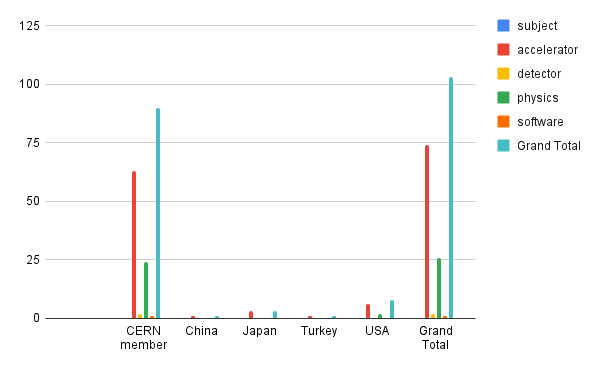}}
\caption{ 
for the   \href{https://indico.cern.ch/event/995850/}{FCC 2021 conference}
[left] talks by country 
[right]   talks by subject
}\label{fig:fcc2021}
\end{figure}

The \href{https://indico.cern.ch/event/995850/}{FCC2021 conference} was held 28 June to 2 July 2021.
Figure~\ref{fig:fcc2021} shows the nation of the speaker's affiliation for all talks (left) and  by several talk categories (physics, accelerator, detector development)(right).

\begin{figure}
\centering
\resizebox{0.48\textwidth}{!}{\includegraphics{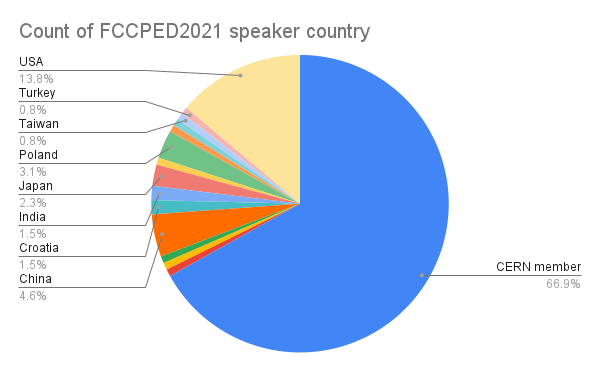}}
\resizebox{0.48\textwidth}{!}{\includegraphics{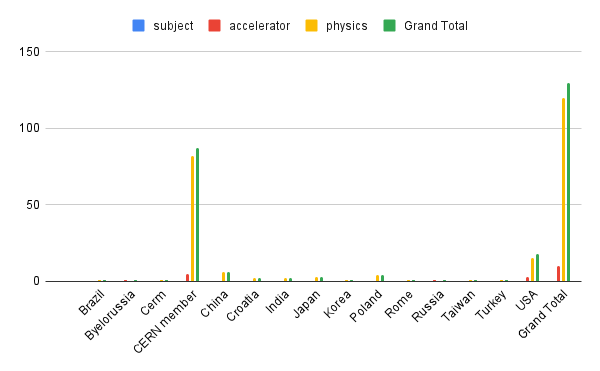}}
\caption{
for the \href{https://indico.cern.ch/event/932973/}{FCC PED 2020 conference}  
[left] talks by country 
[right]   talks by subject
}\label{fig:fccped_2020}
\end{figure}

The \href{https://indico.cern.ch/event/932973/}{FCC ``physics, experiments, and detectors" conference 2020}  was fully online and was help 10-13 November 2020.
Figure~\ref{fig:fccped_2020} shows the nation of the speaker's affiliation for all talks (left) and  by several talk categories (physics, accelerator, detector development) (right).

\begin{figure}
\centering
\resizebox{0.48\textwidth}{!}{\includegraphics{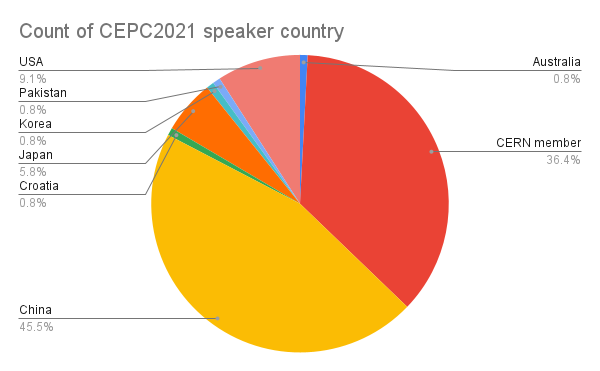}}
\resizebox{0.48\textwidth}{!}{\includegraphics{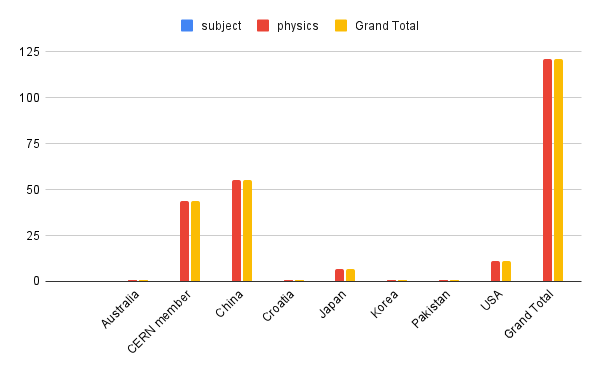}}
\caption{ 
for the \href{https://indico.ihep.ac.cn/event/14938/overview}{CEPC conference} (accelerator excluded due to difficulty parsing data as the titles were in Chinese)  
[left] talks by country 
[right]   talks by subject
}\label{fig:cepc2021}
\end{figure}

The \href{https://indico.ihep.ac.cn/event/14938/overview}{CEPC 2021 conference} was help 8-12 November 2021 and was fully online.
Figure~\ref{fig:cepc2021} shows the nation of the speaker's affiliation for all talks (left) and  by several talk categories (physics, accelerator, detector development) (right).


As can be seen in the distribution of talks, Europeans dominate, while U.S. participation is weaker than it should be to maintain leading roles and influence. Increased support for U.S. efforts would result in stronger presence among the international contributions.

\section{Executive Summary}

It is an exciting time for the study of fundamental physics.  The discovery of the Higgs boson has opened a new window on fundamental physics waiting to be explored.  
More broadly, the U.S. high energy physics program requires success in confronting a number of practical challenges that study of the Higgs boson can address.  These include maintaining leading roles in international partnerships, nurturing and advancing world-leading capabilities and expert resources, and maintaining and attracting talent.  
The global interest in a Higgs factory provides the opportunity to ensure the U.S. program succeeds on these critical issues.  Currently, many countries are providing substantial budgets for Higgs factory experimental preparation.  This sets the scale needed to engage the U.S. community, with its many areas of expertise. With such support, the U.S. capabilities would immediately be welcomed into partnerships with the international effort.  It would sustain the U.S reputation as a partner to be relied on, while building on existing expertise, and retaining and attracting the best young personnel.


\section{Acknowledgments}

We acknowledge support from the following funding agencies: DOE and NSF (USA).

\bibliographystyle{elsarticle-num}
\bibliography{theBIB}

\begin{thebibliography}{10}
\expandafter\ifx\csname url\endcsname\relax
  \def\url#1{\texttt{#1}}\fi
\expandafter\ifx\csname urlprefix\endcsname\relax\def\urlprefix{URL }\fi
\expandafter\ifx\csname href\endcsname\relax
  \def\href#1#2{#2} \def\path#1{#1}\fi

\bibitem{higgscms}
S.~Chatrchyan, et~al.,
  \href{http://dx.doi.org/10.1016/j.physletb.2012.08.021}{Observation of a new
  boson at a mass of 125 {GeV} with the {CMS} experiment at the {LHC}}, Physics
  Letters B 716 (2012) 30–61.
\newblock \href {https://doi.org/10.1016/j.physletb.2012.08.021}
  {\path{doi:10.1016/j.physletb.2012.08.021}}.
\newline\urlprefix\url{http://dx.doi.org/10.1016/j.physletb.2012.08.021}

\bibitem{higgsatlas}
G.~Aad, et~al.,
  \href{http://dx.doi.org/10.1016/j.physletb.2012.08.020}{Observation of a new
  particle in the search for the standard model {Higgs} boson with the {ATLAS}
  detector at the {LHC}}, Physics Letters B 716 (2012) 1–29.
\newblock \href {https://doi.org/10.1016/j.physletb.2012.08.020}
  {\path{doi:10.1016/j.physletb.2012.08.020}}.
\newline\urlprefix\url{http://dx.doi.org/10.1016/j.physletb.2012.08.020}

\bibitem{EuropeanStrag}
{European Strategy Group}, \href{https://cds.cern.ch/record/2721370}{2020
  update of the {European} strategy for particle physics (brochure)},
  CERN-ESU-015 (2020).
\newline\urlprefix\url{https://cds.cern.ch/record/2721370}

\bibitem{mockett}
P.~Mockett, {A review of the physics and technology of high-energy calorimeter
  devices}, Proc. Dyn. Spectrosc. High Energy (1983) 1.

\bibitem{RevModPhys.90.025002}
S.~Lee, M.~Livan, R.~Wigmans,
  \href{https://link.aps.org/doi/10.1103/RevModPhys.90.025002}{Dual-readout
  calorimetry}, Rev. Mod. Phys. 90 (2018) 025002.
\newblock \href {https://doi.org/10.1103/RevModPhys.90.025002}
  {\path{doi:10.1103/RevModPhys.90.025002}}.
\newline\urlprefix\url{https://link.aps.org/doi/10.1103/RevModPhys.90.025002}

\bibitem{Berridge:1988ie}
S.~C. Berridge, P.~K. Berridge, J.~E. Brau, W.~M. Bugg, Y.~Chen, Y.~C. Du,
  R.~S. Kroeger, K.~T. Pitts, A.~W. Weidemann, S.~L. White, {The Small Angle
  Electromagnetic Calorimeter at Sld: A 2-m**2 Application of Silicon Detector
  Diodes}, IEEE Trans. Nucl. Sci. 36 (1989) 339--343.
\newblock \href {https://doi.org/10.1109/23.34460}
  {\path{doi:10.1109/23.34460}}.

\bibitem{ilcx2021}
J.~E. Brau, et~al.,
  \href{https://agenda.linearcollider.org/event/9211/sessions/5248}{{The SiD
  Digital ECal based on Monolithic Active Pixel Sensors }} (2021).
\newline\urlprefix\url{https://agenda.linearcollider.org/event/9211/sessions/5248}

\bibitem{Key4hep}
\href{https://key4hep.github.io/key4hep-doc/}{{Key4hep}}.
\newline\urlprefix\url{https://key4hep.github.io/key4hep-doc/}

\bibitem{miyamoto_akiya_2021_4659567}
A.~Miyamoto, A.~Sailer, F.~Gaede, N.~Graf, J.~Strube, M.~Stanitzki,
  \href{https://doi.org/10.5281/zenodo.4659567}{Computing requirements of the
  ilc experiments} (Apr. 2021).
\newblock \href {https://doi.org/10.5281/zenodo.4659567}
  {\path{doi:10.5281/zenodo.4659567}}.
\newline\urlprefix\url{https://doi.org/10.5281/zenodo.4659567}

\bibitem{MONARC}
\href{https://monarc.web.cern.ch/MONARC/}{{The MONARC Project}}.
\newline\urlprefix\url{https://monarc.web.cern.ch/MONARC/}

\bibitem{osti_1074469}
J.~E. Brau, \href{https://www.osti.gov/biblio/1074469}{Final report for the
  university-based detector research and development for the international
  linear collider} (2013).
\newblock \href {https://doi.org/10.2172/1074469} {\path{doi:10.2172/1074469}}.
\newline\urlprefix\url{https://www.osti.gov/biblio/1074469}

\bibitem{cernmed}
\href{https://cds.cern.ch/record/2735179}{{Medium-Term Plan for the period
  2021-2025 and draft budget of the Organization for the sixty-seventh
  financial year 2021. Video-meeting: Scientific Policy Committee -
  Three-Hundred-and-Nineteenth Meeting}} (Sep 2020).
\newline\urlprefix\url{https://cds.cern.ch/record/2735179}

\bibitem{italypiano}
\href{https://www.presid.infn.it/index.php/it/11-piano-triennale}{{Piano
  Triennale 2021-2023}} (2021).
\newline\urlprefix\url{https://www.presid.infn.it/index.php/it/11-piano-triennale}

\bibitem{bedeschi}
Franco Bedeschi, Private Communication, Jan 2022.

\bibitem{Skorea}
H.~Yoo, \href{https://agenda.infn.it/event/28676/}{{Status of DR work in
  Korea}}, slide 5 (2021).
\newline\urlprefix\url{https://agenda.infn.it/event/28676/}

\end{thebibliography}

\end{document}